\documentstyle[emulateapj] {article}

\begin{document}

\title{The Compact Nucleus of the Deep Silicate Absorption Galaxy NGC
4418\footnote{Some of the data presented herein were obtained at the W. M.
Keck Observatory, which is operated as a scientific partnership among the
California Institute of Technology, the University of California, and the
National Aeronautics and Space Administration.  The Observatory was made
possible by the generous financial support of the W. M.  Keck Foundation.}}

\author{A. S. Evans\altaffilmark{2}, E. E. Becklin\altaffilmark{3},
N. Z. Scoville\altaffilmark{4}, 
G. Neugebauer\altaffilmark{5}, B. T. Soifer\altaffilmark{5},
K. Matthews\altaffilmark{5}, M. Ressler\altaffilmark{6},
M. Werner\altaffilmark{6}, and M. Rieke\altaffilmark{7}} 
\altaffiltext{2}{Dept. of Physics \& Astronomy, Stony Brook University,
Stony Brook, NY 11794-3800: aevans@mail.astro.sunysb.edu}
\altaffiltext{3}{University of California, Los Angeles, CA 90095}
\altaffiltext{4}{Division of Physics, Math, \& Astronomy 105-24,
California Institute of Technology, Pasadena, CA 91125}
\altaffiltext{5}{Palomar Observatory, California Institute of Technology,
320-47, Pasadena, CA 91125}
\altaffiltext{6}{Jet Propulsion Laboratory, 169-506, 4800 Oak Grove Dr.,
Pasadena, CA 91109}
\altaffiltext{7}{Steward Observatory, University of Arizona, Tucson, AZ
85721}

\begin{abstract}

High resolution, Hubble Space Telescope (HST) near-infrared and Keck
mid-infrared images of the heavily extinguished, infrared luminous galaxy
NGC 4418 are presented. These data make it possible to observe the
imbedded near-infrared structure on scales of 10--20 pc, and to constrain
the size of the mid-infrared emitting region.  The $1.1-2.2\mu$m data of
NGC 4418 show no clear evidence of nuclear star clusters or of a reddened
active galactic nucleus. Instead, the nucleus of the galaxy consists of a
$\sim$100--200 pc linear structure with fainter structures extending
radially outward.  The near-infrared colors of the linear feature are
consistent with a 10-300 Myr starburst suffering moderate levels (few
magnitudes) of visual extinction.  At 7.9--24.5$\mu$m, NGC 4418 has
estimated size upper limits in the range of 30--80 pc.  These dimensions
are consistent with the highest resolution radio observations obtained to
date of NGC 4418, as well as the size of 50--70 pc expected for a
blackbody with a temperature derived from the 25$\mu$m, 60$\mu$m, and
100$\mu$m flux densities of the galaxy.  Further, a spectral energy
distribution constructed from the multi-wavelength mid-infrared
observations show the strong silicate absorption feature at 10$\mu$m,
consistent with previous mid-infrared observations of NGC 4418.

An infrared surface brightness of $\sim 2.1\times10^{13}$ L$_\odot$
kpc$^{-2}$ is derived for NGC 4418. Such a value, though consistent with
the surface brightness of warm ultraluminous infrared galaxies (ULIGs:
$L_{\rm IR} [8-1000\micron] \geq 10^{12}$ L$_\odot$) such as IRAS
05189-2524 and IRAS 08572+3915, is not large enough to distinguish NGC
4418 as a galaxy powered by an Active Galactic Nucleus (AGN), as opposed
to a lower surface brightness starburst.

\end{abstract}


\section{Introduction}

Mid-infrared spectra of some infrared luminous galaxies show a deep
10$\mu$m silicate grain absorption feature believed to be a strong
indication of a buried, compact energy source (Roche et al. 1986;
Smith, Aiken, \& Roche 1989; Dudley \& Wynn-Williams 1997; Spoon et
al. 2001).  The nearest of these, the $L_{\rm IR} (8-1000\micron)
\sim 8 \times 10^{10} L_{\odot}$ galaxy NGC 4418 (Roche et al.  1986),
is a nearly edge-on Sa type galaxy which is likely interacting with a
distorted companion galaxy 24 kpc away (see the Digitized Sky Survey
image of the galaxy pair). NGC 4418 has a high infrared
luminosity-to-molecular gas mass ratio ($L_{\rm IR} /$M[H$_2] \gtrsim
100$:  Sanders, Scoville, \& Soifer 1991)\footnote{ For spiral galaxies
and for extreme star-forming giant molecular cloud cores, where stars
are the main source of dust heating,  $L_{\rm IR} /$M(H$_2) = 5$ and
25, respectively; Sanders \& Mirabel (1996).} and warm infrared colors
($f_{25\mu m} / f_{60\mu m} = 0.23$) similar to those observed in infrared
galaxies with Seyfert-like emission line spectra; the latter property
has led to speculation that the dust in NGC 4418 is being heated by an
imbedded AGN.  The optical spectrum of the galaxy has very faint H$\alpha$
emission, but the forbidden lines of [N II] $\lambda\lambda$ 6548,6583,
[O I] $\lambda$ 6300, and [S II] $\lambda\lambda$ 6716,6731 have not been
detected, making it difficult to classify the galaxy as a Seyfert, LINER,
or H II Region-like galaxy (Armus, Heckman, \& Miley 1989).  No emission
lines are detected at mid-infrared wavelengths (Spoon et al. 2001).
The calculated visual extinction of $>$50 mag derived from the
observed silicate absorption feature verifies that optical observations
reveal little about the nuclear regions.

As part of the larger program to image IRAS-luminous galaxies with
the Hubble Space Telescope (HST) Near-Infrared Camera and MultiObject
Spectrometer (NICMOS) and the Keck II Telescope with a mid-infrared camera
(MIRLIN), observations of NGC 4418 have been obtained. Both instruments
are ideally suited for observations of infrared galaxies: NICMOS combines
near-infrared technology with the excellent resolution possible with HST
(0.11\arcsec $~$at 1.1 $\mu$m), making it possible to peer deeper into the
optically obscured nuclear regions of infrared galaxies.  MIRLIN combines
mid-infrared technology with the large collecting area and spatial
resolution capabilities of the Keck telescope.  These observations yield
important insight to the environment of the central energy source of NGC
4418, and possibly to the nature of more distant and infrared luminous
galaxies with deep silicate absorption features.  The NICMOS portion of
this dataset has been briefly discussed elsewhere (Scoville et al. 2000).
Throughout this paper, a distance to NGC 4418 of 27.3 Mpc (e.g. Ridgway,
Wynn-Williams, \& Becklin 1994) is adopted such that 1$\arcsec$ subtends
130 pc at the distance of NGC 4418.

\section{Observations and Data Reduction}

\subsection{Near-Infrared Data}

HST observations of NGC 4418 were obtained on 1997 November 27 using the
NICMOS camera 2. NICMOS consists of three 256$\times$256 HgCdTe arrays,
each with a different pixel scale; the pixel scale of camera 2 is
$0.0762\arcsec\times0.0755\arcsec$ pixel$^{-1}$ in the $x$ and $y$
direction, providing a $19.5\arcsec\times19.3\arcsec$ field of view
(Thompson et al.  1998).  Images were obtained using the F110W (1.1
$\mu$m), F160W (1.6 $\mu$m), and F222M (2.2 $\mu$m) filters, providing
resolutions of 0.11, 0.16, and 0.22\arcsec$~$, respectively.  Observations
were done by executing a 4-point spiral dither per filter setting. At each
dither position, non-destructive reads (MULTIACCUM) were obtained, with
integration times of 88 seconds (1.1 $\mu$m, 1.6 $\mu$m) and 120 seconds
(2.2 $\mu$m) per exposure.  In addition to the galaxy observations,
3-point spiral dither observations of 120 seconds each were done on blank
sky (4$\arcmin$ south of NGC 4418) using the 2.2 $\mu$m filter.  Finally,
dark exposures were taken using the same MULTIACCUM sequences executed for
the galaxy and sky observations.

Reduction of the data was done with IRAF.  The dark was first created,
then the NICMOS data were dark subtracted, flatfielded and corrected for
cosmic rays using the IRAF pipeline reduction routine CALNICA (Bushouse
1997).  The dithered images were then shifted and averaged using the
DRIZZLE routine in IRAF (e.g. Hook \& Fruchter 1997).  The blank sky data
were reduced in the same manner, and the measured 2.22 $\mu$m sky level of
0.45 ADU/sec was subtracted from the 2.2 $\mu$m image of NGC 4418.
Additional sky subtraction was performed for each band by removing an
offset equal to the average pixel value near the edges of each image. Flux
calibrations of the images were calculated using the conversion factors
2.031$\times10^{-6}$, 2.190$\times10^{-6}$, and 5.487$\times10^{-6}$~Jy
(ADU/sec)$^{-1}$ at 1.1, 1.6, and 2.22 $\mu$m, respectively, and
corresponding magnitude zero-points of 1775, 1083, and 668 Jy.

To enhance the details seen in all three bands and to achieve a common
resolution, all three images were deconvolved using the IRAF routine LUCY
and a Point-Spread Function (PSF) star observed on 1997 November 20 with
the same filters and dither pattern as the NGC 4418 observations.  The 1.1
$\mu$m and 1.6 $\mu$m images were then convolved with a Gaussian using the
routine GAUSS such that their resolutions matched that of the deconvolved
2.2 $\mu$m image ($\sim 0.14\arcsec$)\footnote{This was achieved by first
performing the same analysis on infrared galaxy images containing
unresolved star clusters (see Scoville et al. 2000); the quoted resolution
is the resolution of these star clusters after the analysis.  Gaussians
with different dispersions were applied to these images until one with the
appropriate resolution was achieved.}.  The resultant images are shown in
Figures 1--2; Figure 1 shows the 1.1 $\mu$m, 1.6 $\mu$m, 2.2 $\mu$m, and
$1.6 \mu$m $/ 1.1 \mu$m images separately in greyscale, and Figure 2 is a
three-color image of NGC 4418.

\subsection{Mid-Infrared Data}

The mid-infrared observations of NGC 4418 were made on 1998 March 19 using
the MIRLIN mid-infrared camera (Ressler et al. 1994) at the f/40 bent
Cassegrain visitor port of the Keck II Telescope. The camera uses a
128$\times$128 Si:As array with a plate scale of 0.138$''$/pixel for a
total field of view of 17$'' \times$17$''$.  Table~1 gives the central
wavelengths and filter bandwidths for the filters through which the object
was observed.  At each wavelength the observing procedure was the same.  A
secondary mirror with a square wave chop of  amplitude 6$''$ in the east-west
direction at 4 Hz was employed for fast beam switching.  The frames
sampling  each chop position were coadded separately in hardware,
resulting in two images.  After an interval of approximately a minute, the
telescope was nodded perpendicular to the chop direction  by 6$''$ and a
second pair of images was obtained in order to cancel residuals in the sky
and to subtract telescope emission.  This procedure was repeated a number
of times at each  wavelength.  The data were reduced by differencing the
two images obtained within the chop pairs at each nod location, and then
coadding the resulting positive images, with the positions appropriately
adjusted to a common location, to yield a positive image centered in a
field approximately 6$''~\times$~6$''$.  Because of the  chopper and
telescope nod spacings employed for the observations, the  data are not
capable of measuring flux outside a 6$''$ diameter region.

The main objective of the observational program was focused on obtaining
mid-infrared photometry of a large number of sources. Hence only limited
observations were made to constrain the size of NGC 4418.  The
observations of NGC 4418 were immediately followed by observations of Mrk
231 and  HR 5340 both of which are assumed to be point-like, i.e., the
objects are assumed to be PSFs. These observations took nearly an hour. 
At 24.5~$\mu$m the PSF's throughout the night equaled the diffraction limit
of the telescope; i.e., 0.62$''$ FWHM.  At the shorter wavelengths,
the size of the PSF image was set by diffraction and the pixel sampling
about half of the time throughout the night; at other times atmospheric
seeing affected the image size. The variation in the measured size of
the PSF calibration star is  the largest uncertainty in the limit of
the source size.

\section{Photometry}

\subsection{Near-Infrared}

Carico et al. (1988) measured $J, H,$ and $K$ magnitudes of NGC 4418 of
12.97, 12.20, and 11.85, respectively, in a 5$\arcsec$-diameter beam.
The corresponding $J-H$ and $H-K$ colors are 0.77 and 0.35, respectively.
For comparison, magnitudes for the three NICMOS images were calculated
using a 5$\arcsec$ beam, and are measured to be $m_{1.1 \mu m} =
13.22$ mag, $m_{1.6 \mu m} = 12.21$ mag, and $m_{2.2 \mu m} = 11.77$ mag
(see also Table 2); the latter two measurements are within 0.1 mag
of the H and K measurements by Carico et al. (1988).  Interpolating the
1.25$\mu$m magnitude from the 1.1 and 1.6 $\mu$m magnitudes, $m_{1.25 \mu
m}$ is determined to be 12.92 mag, which is also within 0.1 mag of the
Carico et al. $J-$band measurement.  Thus, $m_{1.1 \mu m - 1.6 \mu m} =
1.00$ mag, $m_{1.25 \mu m- 1.6 \mu m} = 0.71$ mag, and  $m_{1.6 \mu m -
2.2 \mu m} = 0.44$ mag.  Subtracting the underlying galaxy from the
nuclear component (within $1\farcs1$, which is large enough to enclose the
first Airy ring of a 2.2 $\mu$m NICMOS PSF) yields $m_{1.1 \mu m - 1.6 \mu
m} = 1.26$ mag, $m_{1.25 \mu m - 1.6 \mu m} =0.88$ mag, and $m_{1.6 \mu m
- 2.2 \mu m}=0.73$ mag.

\subsection{Mid-Infrared}

The mid-infrared photometric data are summarized in Table 1.  The
photometry  was calibrated based on observations of the bright star
HR 1457 (= $\alpha$ Tau) whose magnitudes, in turn, were based on IRAS
photometry and intercomparisons with other bright stars over several
nights. Only the statistical uncertainty is given in Table~1. The
uncertainties in the photometry, based on the internal consistency of
the observations, are estimated to be 5$\%$ at $\lambda \le$17.9~$\mu$m
and 10\% at 24.5~$\mu$m.  The flux density corresponding to 0.0 mag
(Vega-based)  was taken to follow the prescription given in the
Explanatory Supplement to the IRAS Catalogs and Atlases (Beichman et
al\ 1989).

\section{General Characteristics of the Nucleus}

\subsection{Near-Infrared}

Figure 1a--c show greyscale images of the nuclear region at 1.1, 1.6, and
2.2 $\mu$m.  The inner 300 pc of NGC 4418 has a spider-like morphology:
The highest surface brightness emission emanates from a central linear
structure $\sim$100-200 pc across at a position angle of $\sim 110^{\rm
o}$.  Fainter emission extends radially outward from the linear structure,
with the highest surface brightness extension oriented perpendicular to
the linear structure.  The resultant ``T''-like appearance of the higher
surface brightness nuclear features is most apparent at 1.1 $\mu$m.

Figure 3 shows plots of the intensity profiles along the major axis of the
linear structure. The intensity profile shows marked asymmetries which may
be due to extinction by foreground dust. The inner $\sim 3-4$ pixels of
each image are within a resolution element of the images -- outside of
this range, the intensity profiles initially decrease in a nearly gaussian
manner with radius, then exponentially. The width of the nucleus also
decreases with wavelength; the full width at half the maximum intensity is
0.80$\arcsec$ (105 pc) at 1.1 $\mu$m and 0.38$\arcsec$ (50 pc) at 2.2
$\mu$m. Beyond a full width of 1.82$\arcsec$ (237 pc), the slopes of the
profiles are shallow, and the intensity is most likely dominated by
emission from the underlying galaxy.

In addition to the radial emission structures, two dark lanes extend
radially southwest from the nucleus (see Figure 2). A more diffuse dark
lane extends in the northern direction.

\subsection{Mid-Infrared}

By a comparison with the sizes measured on HR 5340 and Mrk 231, NGC 4418
is unresolved at most mid-infrared wavelengths with these data. The upper
limit depends on the wavelength, but subtracting the size of HR 5340 or
Mrk 231 from NGC 4418 in quadrature yields a size of approximately
0.23$\arcsec$ (30 pc) diameter at 8 $\mu$m, and there is perhaps some
evidence that it is marginally extended by around 0.31$\arcsec$ (40 pc)
diameter in the silicate absorption feature near 10 $\mu$m.  At 24.5
$\mu$m, the upper limit on the size is 0.61$\arcsec$ (80 pc).

\section{Discussion}

These new near- and mid-infrared data provide a significant improvement in
resolution relative to previously published data at these wavelengths;
near-infrared images of NGC 4418 obtained by Zenner \& Lenzen (1993) have
a resolution between 0.5$\arcsec$ and 1.3$\arcsec$ , and previously
published 7--15 $\mu$m (Dale et al. 2000) and 12--25$\mu$m (Wynn-Williams
\& Becklin 1993) data have resolutions of $\sim 8\arcsec$ and 3$\arcsec$,
respectively.

The near-infrared images of NGC 4418 are both striking and bemusing.  In
contrast to other luminous infrared galaxies imaged by NICMOS as part of
the NICMOS GTO program (Scoville et al. 2000), the nucleus of NGC 4418
shows no direct evidence of star clusters. Nor is there evidence of an
unresolved AGN as has been observed for warm luminous infrared galaxies
such as NGC 7469 (e.g. Scoville et al. 2000).  Instead, it consists of a
resolved, central high surface brightness structure with light and dark
features extending radially outward.

As discussed earlier, NGC 4418 has mid-infrared colors consistent with
infrared galaxies observed to have Seyfert-like emission line spectra.
However, unlike warm galaxies such as NGC 7469, the near-infrared colors
of nucleus are not consistent with reddened AGN-light (see Figure 5 of
Scoville et al. 2000). If the near-infrared emission is stellar in origin,
then the age of the nuclear stellar population can be constrained by
comparing its NICMOS colors with stellar population synthesis models in
which the stellar light has been matched with the bandpasses of the NICMOS
filters. Models have been derived for a 3 -- 300 Myr Bruzual \& Charlot
(1993; 1995; private communications) instantaneous starburst population
having {\it (i)} a Salpeter initial mass function {\it (ii)} a mass range
0.1--125 M$_\odot$, {\it (iii)} and solar metallicity.  A comparison
between these models and the nuclear colors of NGC 4418 shows the nuclear
colors to be redder than the models.  The application of the Rieke \&
Lebofsky (1985) extinction law show the nuclear colors to be consistent
with a 10-300 Myr starburst suffering 2 (screen model) to 5 (dust mixed
with stars model) magnitudes of visual extinction (see also Scoville et
al.  2000).  Near-infrared spectroscopy of NGC 4418 (Ridgway,
Wynn-Williams, \& Becklin 1994) shows a 2.3 $\mu$m CO absorption feature
(in a 2.7$\arcsec$ diameter beam centered on the nucleus); the relative
depth of the CO absorption feature and the slope of the continuum emission
are consistent with a late-type supergiant or metal-rich giant
population.  If the near-infrared light from the nuclear feature is
supergiant starlight with a few magnitudes of visual extinction, this
would indicate that the supergiants responsible for the 2.3$\mu$m
absorption feature are in front of the bulk of the silicate grains
responsible for the absorption feature of NGC 4418 at 9.7$\mu$m (Roche et
al. 1986; Spoon et al. 2001).

\subsection{Near- to Far-Infrared Spectral Energy Distribution}

Figure 4 is a spectral energy distribution (SED) of NGC 4418 using the
photometry derived from the NICMOS and MIRLIN observations, combined
with photometry from other sources (Carico et al.  1990; IRAS FSC 1990;
Dale et al. 2000); the beam sizes are listed in the Figure legend. For
comparison, the ISO-PHT-S data from Spoon et al. (2001) are also plotted.
Three key features of this SED are {\it (i)} a near-infrared thermal
component which is likely due to the late-type supergiant population,
{\it (ii)} the shape of the MIRLIN SED at mid-infrared wavelengths which
clearly shows the presence of the silicate absorption feature in NGC 4418
observed by Roche et al.  (1986) and Spoon et al.  (2001) and {\it (iii)}
the size of the 12 and 25$\mu$m flux densities measured by IRAS which are
recovered by the unresolved MIRLIN emission at those wavelengths. From
the latter two features, it can be concluded that the energy source(s)
heating the silicate grains are confined to an area no more than 80 pc
($\sim 0.6\arcsec$) across.  Further, the highest resolution radio map
of NGC 4418 obtained to date shows the radio emission to be less than
$0.47\arcsec$ in extent (Eales et al. 1990), providing further proof
of the importance of the inner 80 pc to the total energy output of
the galaxy.

\subsection{Far-Infrared Size Constraint and Surface Brightness}

Even given the overwhelming evidence at mid-infrared and radio wavelengths
that the energy sources in NGC 4418 are confined in a nuclear region no
more that 80 pc across, a direct measurement at the wavelength where most
of the energy of NGC 4418 is emanating from -- the far-infrared -- is
ideally desired.  High-resolution imaging of the galaxy at these
wavelengths is not presently possible, however, additional support for a
compact far-infrared emission region can be provided via two
approximations.

For the first approximation, the assumption is made that the nuclear dust
that is re-radiating light from the imbedded nuclear source(s) in NGC
4418 is distributed in an optically thick sphere of diameter $D$ and
outer blackbody temperature, $T_{\rm dust}$. The dust temperature is 
calculated from the 60 and 100 $\micron$ flux densities, $f_{60\mu{\rm m}}$
and $f_{100\mu{\rm m}}$, via the equation

$$T_{\rm dust} = -(1+z) \left[{82 \over \ln (0.3f_{60\mu{\rm
m}}/f_{100\mu{\rm m}})} - 0.5 \right] \eqno(1)$$

\noindent
(e.g., see Solomon et al. 1997).  For $f_{60\mu{\rm m}} \sim 40.68$ Jy and
$f_{100\mu{\rm m}} \sim 32.80$ Jy, $T_{\rm dust} \sim 85$ K. The blackbody
diameter is thus calculated via the equation

$$D = 2 \left[ {L_{\rm IR} \over 4 \pi \sigma T^4_{\rm dust}} \right]
^{0.5}, \eqno(2)$$

\noindent
where $\sigma$ is the Stefan-Boltzmann constant, which yields a diameter
of 0.54$\arcsec$ (70 pc). This is equivalent to the upper size limit of
the 25$\mu$m and radio emission of NGC 4418, as well as the FWHM of its
2.2$\mu$m emission. Note that a source with a brightness temperature
of 85K and a flux density of 9.32 Jy at 25 $\mu$m has a diameter of
0.41$\arcsec$ (53 pc), which is also consistent with the measured upper
limit of the 25 $\mu$m emission.

If the dust emitting the far-infrared emission is associated with the
star-forming molecular gas in NGC 4418, the distribution of this gas in
the nucleus of NGC 4418 can be used as an independent second approximation
to the extent of the far-infrared emission. No interferometric CO($1\to0$)
data presently exist of NGC 4418, however single-dish measurements have
been made which provide a CO luminosity, $L'_{\rm CO}$, and a velocity
dispersion.  Using these data, the assumption is made that the brightness
temperature of the CO($1\to0$) emission is equal to the dust temperature
and that the gas has a unity filling factor. Thus, the CO diameter,
$D_{\rm CO}$, is derived via

$$D_{\rm CO} = 2 \sqrt{L'_{\rm CO} \over \pi {T_{\rm bb} \Delta v_{\rm
FWHM}} }, \eqno(3)$$

\noindent
where $\Delta v_{\rm FWHM}$ is the full width at half the maximum
intensity velocity width (= 120 km s$^{-1}$: Sanders, Scoville, \& Soifer
1991). The CO diameter is calculated to be 1.5$\arcsec$ (200 pc).

If the size of 70 pc is adopted for the mid-to-far infrared emission
region, then the surface brightness of NGC 4418 is calculated to be $4
L_{\rm IR} / \pi D^2 \sim 2.1\times10^{13}$ L$_\odot$ kpc$^2$. Figure
5 is a plot of infrared surface brightness versus infrared luminosity
for a sample of galaxies imaged at mid-infrared wavelengths by Soifer
et al. (2000; 2001), as well as the Orion star-forming complex and M82
(see also Soifer et al.  2001). The surface brightness of NGC 4418 is
extreme, being comparable to that of the warm ultraluminous infrared
galaxies (e.g., IRAS 05189-2524, IRAS 08572+3915, and Mrk 231) and
cool ultraluminous infrared galaxies ($f_{25\mu m} / f_{60\mu m} <
0.2$; which may be powered primarily by starburst, AGN, or both) such
as Arp 220, UGC 5101, and Mrk 273, a factor of 100 larger than that of
M82, and a factor of ten larger than that of the star forming complex in
Orion.\footnote{Note that, with the exception of Arp 220, UGC 5101, and IR
17208-0018, all of the ultraluminous infrared galaxies plotted in Figure 5
show strong evidence of being powered by an AGN (i.e., based on the
presence of broad recombination emission lines, warm mid-infrared colors,
and/or AGN-like X-ray emission).} In comparison to starburst galaxies with
infrared luminosities moderately higher than NGC 4418 (i.e., NGC 1614, NGC
2623, IC 883, NGC 6090, Mrk 331: Soifer et al. 2001), NGC 4418 has an
infrared surface brightness a factor of 2--100 higher.  From the limited
number of objects plotted in Figure 5, there is no clear indication that
infrared surface brightness alone can be used as a diagnostic between AGN
and starburst energy sources.

\subsection{Dark Lanes}

As previously mentioned, dark lanes are observed to be associated with
the central high surface brightness near-infrared peak of NGC 4418.
Figure 1d shows an $m_{1.1\mu{\rm m}}$ - $m_{1.6\mu{\rm m}}$ image
of NGC 4418; in this image, the dark lanes appear as cones extending away
from the nucleus.

There are two likely explanations for the nature of the dark lanes in the
nucleus of NGC 4418. The first possibility is that the features are dust
lanes.  Such radial dust lanes are observed in the nucleus of spiral
galaxy M51, albeit on a smaller scale (50 pc:  Grillmair 1997 and
references therein).  The presence of such dust lanes, if they lie in the
foreground of the nuclear infrared disk, may explain the asymmetries
observed in Figure 3, however, it is difficult to understand why such
radial dust lanes have not been destroyed by rotation.

The second possible explanation for the dark lanes is that they are
shadows cast from a partially obscured AGN. Such a possibility would
require that the AGN be visible along lines of sight that lie within the
plane of the sky, but also require that certain lines of sight toward the
AGN in the plane of the sky are blocked, perhaps by the presence of
molecular clouds.

These two possibilities can be tested. First, if AGN light is visible
along lines of sights within the plane of the sky, then near-infrared,
low surface-brightness light surrounding the
inner nucleus should be polarized. Second, if the dark lanes are nuclear
dust lanes, CO($1\to0$) observations of NGC 4418 should reveal evidence of
molecular gas coindent with the lanes.  Such CO observations would be
useful in their own right as a confirmation of the high extinction to the
nucleus calculated via the silicate absorption feature.

\subsection{A Portrait of NGC 4418}

Given the new data presented in this paper and compiled from the
literature, NGC 4418 appears to be a galaxy with nuclear power source(s)
which has(have) been triggered via interactions with a companion galaxy 24
kpc away. The galaxy NGC 4418 has the following near- to mid-infrared
properties:

\noindent
{\it (i)} A compact, near-infrared nucleus consisting of a 100--200 pc
high-surface brightness linear feature surrounded by radial extensions,
and near-infrared colors consistent with moderately extinguished
supergiant stellar light,

\noindent
{\it (ii)} No nuclear star clusters at near-infrared wavelengths,
which are commonly seen in starburst infrared galaxies, and no
bright point-like nucleus commonly seen in AGN,

\noindent
{\it (iii)} An infrared surface brightness comparable to warm
ultraluminous infrared galaxies and cool ultraluminous infrared
galaxies such as Arp 220 and UGC 5101,

\noindent
{\it (iv)} Emission of most of its mid to far-infrared light from a region
$\lesssim 80$ pc across,

\noindent
{\it (v)} A deep silicate absorption feature, which translates into $>50$
mag of visual extinction to the central energy source(s) (see
also Roche et al. 1986; Dudley \& Wynn-Williams 1997; Spoon et al. 2001),

\noindent
{\it (vi)} No evidence of strong PAH features, which are common to
mildly extinguished starburst galaxies like M 82 (Genzel et al. 1998;
Spoon et al. 2001), and

\noindent
{\it (vii)} Warm infrared colors ($f_{25\mu m} / f_{60\mu m} = 0.23$)
consistent with those observed for infrared galaxies with Seyfert-like
emission line spectra.

The basic picture of NGC 4418 is one in which a stellar core/disk
approximately 150 pc in extent surrounds an imbedded AGN or a compact
starburst. So deeply buried is the central engine(s) that even probing the
galaxy at near-infrared wavelengths, where extinction is 5--10 mag
rather than 50--100 mag at optical wavelength, provides little
information about the primary energy source in NGC 4418. Observations at
longer wavelengths, using instruments such as MIRLIN, Satellite InfraRed
Telescope Facility (SIRTF), and the Stratospheric Observatory For Infrared
Astronomy (SOFIA), will be required to determine the fundamental nature
of this galaxy class.

\acknowledgements
We thank B. Stobie, J. Mazzarella, D. Dale, A. Sargent, and L. Armus for
useful discussions and assistance, and the anonymous referee for a careful
reading of the manuscript. ASE also thanks H. Spoon for providing
ISO-PHT-S data for inclusion in Figure 4.  ASE and NZS were supported by
NASA grant NAG 5-3042. ASE was also supported by NSF grant AST 02-06262.
This research has made use of the NASA/IPAC Extragalactic Database (NED)
which is operated by the Jet Propulsion Laboratory.


\begin{deluxetable}{cccccc}
\tablewidth{0pt}
\tablenum{1}
\tablecaption{Mid-Infrared Photometry}
\tablehead{
\colhead{Name}&\colhead{Central}&\colhead{Width}&\colhead{HR~1457}&
\colhead{NGC~4418}&\colhead{NGC~4418}\\
\colhead{}&\colhead{Wavelength}&\colhead{}&
\colhead{Magnitude\tablenotemark{a}}&
\colhead{Magnitude\tablenotemark{b}}&
\colhead{Flux Density\tablenotemark{b,c}} \\
\colhead{}&\colhead{$\mu$m}&\colhead{$\mu$m}&\colhead{mag}&
\colhead{mag}&\colhead{mJy}
}
\startdata
 7.9 &  7.91 & 0.76 & -2.99 & 4.85 $\pm$ 0.07 &  704 \\
 8.8 &  8.81 & 0.87 & -3.01 & 5.62 $\pm$  0.04 &  291 \\
 9.7 &  9.69 & 0.93 & -3.03 & 5.98 $\pm$  0.12 &  169 \\
10.3 & 10.27 & 1.01 & -3.04 & 6.60 $\pm$  0.10 &   86 \\
11.7 & 11.70 & 1.11 & -3.06 & 5.27 $\pm$  0.07 &  231 \\
12.5 & 12.49 & 1.16 & -3.07 & 3.51 $\pm$  0.02 & 1030 \\
17.9 & 17.90 & 2.00 & -3.05 & 2.10 $\pm$  0.04 & 1810 \\
24.5 & 24.48 & 0.76 & -3.03 & -0.18 $\pm$ 0.03 & 7670 \\
\enddata
\tablenotetext{a}{adopted}
\tablenotetext{b}{4$''$ diameter beam}
\tablenotetext{c}{f$_\nu$ based on IRAS formulation for 0.0 mag
(Beichman et al.\ 1989)}
\end{deluxetable}

\begin{deluxetable}{cccc}
\tablewidth{0pt}
\tablenum{2}
\tablecaption{HST NICMOS Photometry$^{\rm a}$}
\tablehead{
\multicolumn{1}{c}{Wavelength} &
\multicolumn{1}{c}{Nuclear$^{\rm b}$} &
\multicolumn{1}{c}{1.1$\arcsec$} &
\multicolumn{1}{c}{5$\arcsec$} \nl
\multicolumn{1}{c}{($\mu$m)} &
\multicolumn{1}{c}{mag} &
\multicolumn{1}{c}{mag} &
\multicolumn{1}{c}{mag}
}
\startdata
1.1 & 16.04 & 14.98 & 13.22 \nl
1.6 & 14.78 & 13.85 & 12.21 \nl
2.2 & 14.05 & 13.26 & 11.77 \nl
\enddata
\tablenotetext{a}{These measurements have also appeared in 
Scoville et al. (2000).}
\tablenotetext{b}{Measured in a 1.1$\arcsec$ diameter beam with the
underlying galaxy subtracted.}
\end{deluxetable}

 
\centerline{Figure Captions}
 
\vskip 0.3in
 
\noindent
Figure 1. The deconvolved/Gaussian smoothed images of nuclear regions of
NGC 4418 at (a) 1.1 $\mu$m, (b) 1.6 $\mu$m, (c) 2.2 $\mu$m. The resolution
of each image is 0.14$\arcsec$, and the associated scale bars are in units
of $\mu$Jy pixel$^{-1}$.  Note the morphological similarities of the three
images.  (d)  A $m_{1.1 \mu{\rm m}} - m_{1.6 \mu{\rm m}}$ image of the
nuclear region of NGC 4418.  The green lines outline the edges of the
structure of the dark lanes visible in Figures 1a--c and Figure 2, and the
scale bar represents the ratio of the 1.6 $\mu$m and 1.1 $\mu$m flux
densities provided in (a) and (b).


\noindent
Figure 2.  A composite of three infrared maps of NGC 4418 shown in false
colors. The 1.1 $\mu$m, 1.6 $\mu$m, 2.2 $\mu$m images are shown as blue,
green, and red, respectively.  The field of view is $\sim 11.4\arcsec
\times 11.4\arcsec$.  The logarithm of each image was taken before
combining the three wavelengths to compress the dynamic range.

\noindent
Figure 3. Intensity profile plots of the nuclear structure. The profiles,
which have a width of 0.0386$\arcsec$, have been extracted from the
deconvolved/Gaussian smoothed images.

\noindent
Figure 4. Spectral energy distribution of NGC 4418 based on broad-band
measurements. The solid line is a fit to the broad-band data. Note the
agreement between the broad-band data and the ISO-PHT-S data.

\noindent
Figure 5. A plot of the infrared surface brightness versus infrared 
luminosity for NGC 4418 and a sample of objects discussed in Soifer et
al. (2000; 2001).  The dashed curve is an example of an iso-surface area
curve; all such curves on Figure 5 are parallel to the dashed curve. 

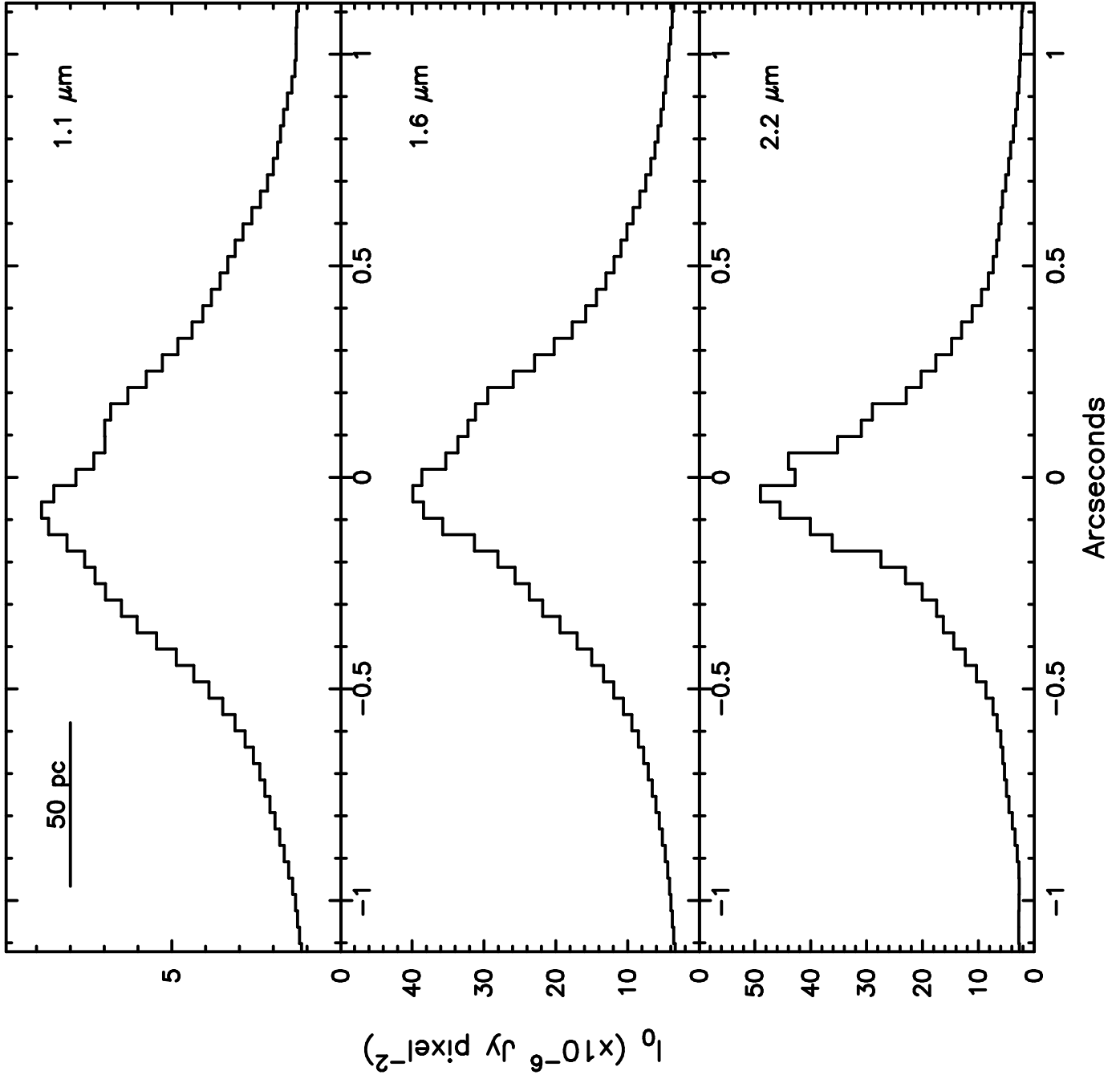
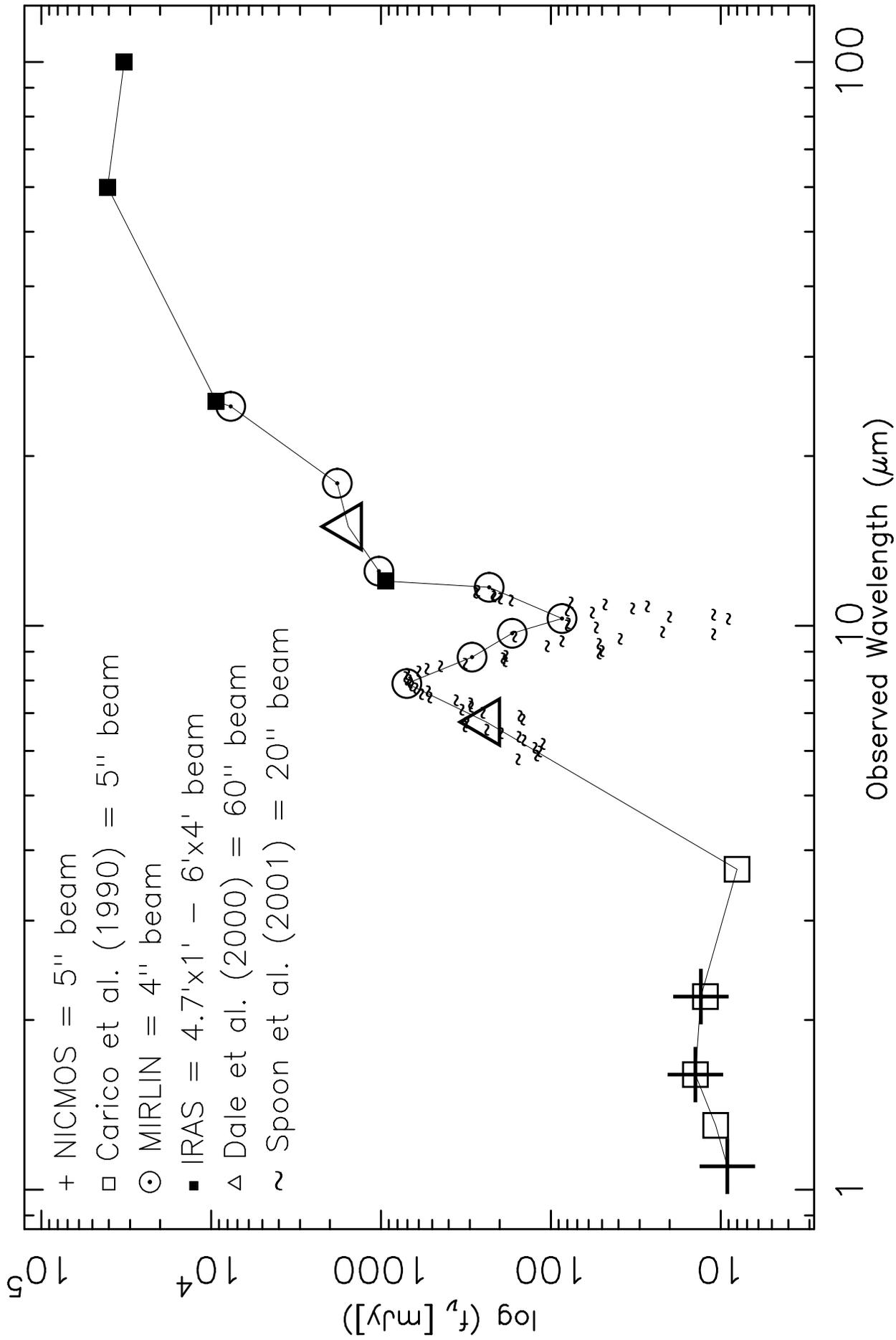
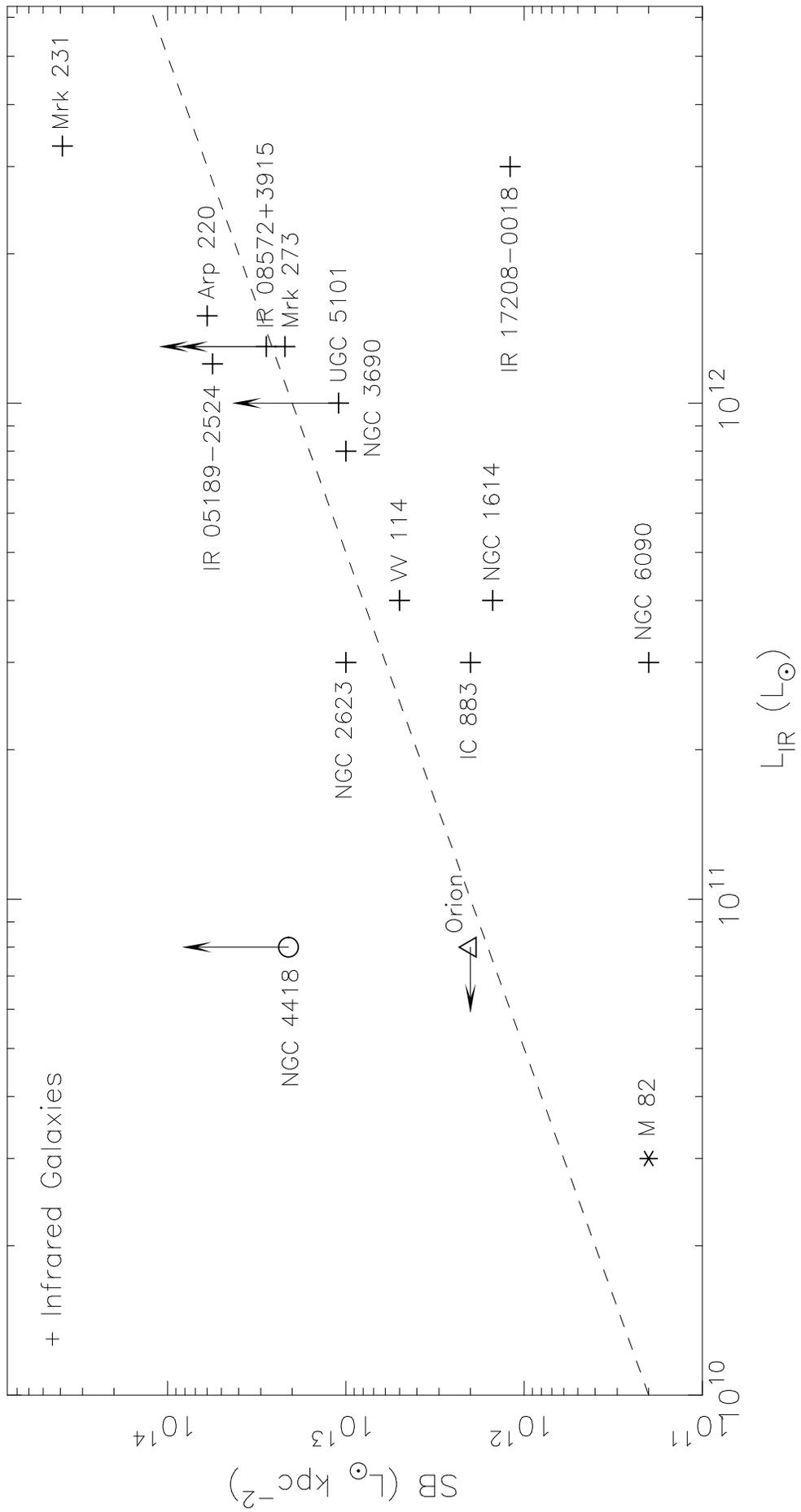
\end{document}